\documentclass[12pt,aps,floats,showpacs,amssymb,tightenlines]{revtex4}

\usepackage{color}
\usepackage{graphicx}
\usepackage{amsmath}
\usepackage{amsfonts}
\usepackage{amscd}
\usepackage{epsfig}
\usepackage{amssymb}
\usepackage{tabularx}

\begin{document}

\title{Axial gravitational perturbations of an infinite static line source.}
\author{Reinaldo J. Gleiser} \email{gleiser@fis.uncor.edu}

\affiliation{Instituto de F\'{\i}sica Enrique Gaviola and FAMAF,
Universidad Nacional de C\'ordoba, Ciudad Universitaria, (5000)
C\'ordoba, Argentina}

\begin{abstract}
In this paper we study axial gravitational perturbations of an infinite static line source, represented by a form of the Levi-Civita metric. The perturbations are restricted to axial symmetry but break the cylindrical symmetry of the background metric. We analyze the gauge issues that arise in setting up the appropriate form of the perturbed metric and show that it is possible to restrict to diagonal terms, but that this does not fix the gauge completely. We derive the perturbation equations and show that they can be solved by solving a third order ordinary differential equation for an appropriately chosen function of the perturbed metric coefficients. The set of solutions of this equation contains gauge trivial parts, and we show how to extract the gauge non trivial components. We introduce appropriate boundary conditions on the solutions and show that these lead to a boundary value problem that determines the allowed functional forms of the perturbation modes. The associated eigenvalues determine a sort of ``dispersion relation'' for the frequencies and corresponding ``wave vector'' components. The central result of this analysis is that the spectrum of allowed frequencies contains one unstable (imaginary frequency) mode for every possible choice of the background metric. The completeness of the mode expansion in relation to the initial value problem and to the gauge problem is discussed in detail, and we show that the perturbations contain an unstable component for generic initial data, and, therefore, that the Levi-Civita space times are gravitationally unstable. We also include, for completeness, a set of approximate eigenvalues, and examples of the functional form of the solutions.

\end{abstract}


\maketitle
\section{Introduction}

In this paper we study axial gravitational perturbations of an infinite static line source, represented by a form of the Levi-Civita metric \cite{levi}, given by Thorne \cite{thorne}. This metric has been used and studied in many instances, including also the possibility of time dependence, but keeping the cylindrical symmetry of the metric. In this case the gravitational perturbations depend only on time and the radial coordinate, and correspond to a form of Einstein - Rosen waves \cite{Einros}. Analysis where the cylindrical symmetry is broken have appeared only recently \cite{konkowski}, but restricted to the evolution of scalar, or Maxwell test fields, in the field of the static line source. The main scope of these studies was directed at elucidating the relation between the classical and quantum nature of the naked singularity present in the background metric. Our aim here is different. What we want to study is if, given the presence of the naked singularity, even after imposing appropriate boundary conditions, the gravitational perturbations contain an unstable component, as has previously been shown to happen for the negative mass Schwarzschild \cite{Schw}, and super extreme Reissner-Nordstrom and Kerr black holes\cite{RN}. As is shown here, we also find in this case unstable modes. We remark again that, as is the case in both \cite{Schw}, and \cite{RN}, the instability is a consequence of the shape of a related ``potential'' away from the singularity, and, therefore, it is possible that it remains even after the (curvature) singularity is smoothed out by spreading the source over a small region. The study of that possibility is, however, outside the scope of the present research.

As indicated, the perturbations we consider here are restricted to axial symmetry but break the cylindrical symmetry of the background metric. In Section III we analyze the gauge issues that arise in setting up the appropriate form of the perturbed metric and show that it is possible to restrict to diagonal terms, but that this does not fix the gauge completely. We analyze in detail the gauge ambiguity remaining after imposing the diagonal form, and show that some of this may be removed by restricting to perturbations that are effectively of compact support. We find several gauge invariants that are used later in the analysis. We derive in the perturbation equations in Section IV, and in Section V we show that they can be solved by solving a third order ordinary differential equation for an appropriately chosen function of the perturbed metric coefficients. The set of solutions of this equation contains gauge trivial parts, and we show how to extract the gauge non trivial components. We introduce appropriate boundary conditions on the solutions in Section VI and show that these lead to a boundary value problem that determines the allowed functional forms of the perturbation modes. The corresponding eigenvalues determine a sort of ``dispersion relation'' for the frequencies and corresponding ``wave vector'' components. Since we not have analytic solutions we use a numerical approach to determine approximate values for eigenvalues and approximate functional form of the eigenfunctions, for a set of background metrics. The central result of this analysis is that the resulting spectrum of allowed frequencies contains one unstable (imaginary frequency) mode for every possible choice of the background metric. The completeness of the mode expansion in relation to the initial value problem and the gauge problem is discussed in detail in Section VII, and we show that the perturbations contain an unstable component for generic initial data. This result implies that the Levi-Civita space times are gravitationally unstable. We close the paper in Section VIII with some final comments.

\section{The metric of a static solution line source.}

The metric of a static solution line source may be written in the form \cite{thorne},
\begin{equation}\label{ATeq1}
ds^2 = e^{2\gamma-2 \psi}
\left(dr^2-dt^2\right)+ e^{2 \psi}dz^2 +e^{-2
\psi}r^2 d\phi^2
\end{equation}
where in general we have:
\begin{eqnarray}\label{ATeq2}
    \psi(r) & = & - \kappa \ln(r/R_0) \nonumber \\
    \gamma(r) & = & \gamma_0+ \kappa^2 \ln(r/R_0)
\end{eqnarray}
with $\gamma_0$, and $R_0$ arbitrary constants that can be reset by changing the scales of $t,r,z,\phi$, while $\kappa$ is a positive constant related to the mass per unit length of the line source. Without loss of generality and mostly for simplicity we will set $\gamma_0=0$ and $R_0=1$ in the rest of this paper. The metric (\ref{ATeq1}) may be thought of as the vacuum metric outside of a massive cylinder, in the limit where the radius of the cylinder goes to zero, but the mass per unit length is kept fixed at some finite value. Notice that in the original form of the metric, analyzed, for instance in \cite{bonnor}, the metric contains a parameter $\sigma$, for which the line source interpretation is possible only for $0\leq \sigma < 1/2$. One can check that this range corresponds to $0 \leq \kappa < \infty$.

\section{Perturbations along the symmetry axis I. The gauge problem.}

In this paper we consider perturbations that preserve the axial symmetry. These correspond to terms in the metric that depend only on $t$, $r$, and $z$. If we call $g^{0}_{ab}$ the static metric given by (\ref{ATeq1}) and (\ref{ATeq2}), and $g^{1}_{ab}$ the perturbation, the perturbed metric $g$ is given by,
\begin{equation}\label{equ01}
    g_{ab} = g^{0}_{ab} +\epsilon g^{1}_{ab}
\end{equation}
where $\epsilon$ is the perturbation parameter, and we will consider all expression only up to first order in $\epsilon$.

We restrict the perturbations $g^{1}_{ab}$ to axial symmetry, and set $g^{1}_{ab}=0$ if one (but not both) of the indices corresponds to $\phi$. This implies, in principle that we have seven independent functions for $g^{1}_{ab}$, but we may use the freedom to change the $(t,r,z)$ coordinates also to first order in $\epsilon$ to restrict to only diagonal terms. In more detail, we consider a (linearized) change of coordinates change of the form,
\begin{eqnarray}
\label{gp02}
  t &=& T+\epsilon t_1(\rho,T,Z) \nonumber \\
  r &=& \rho+\epsilon r_1(\rho,T,Z)   \\
  z &=& Z+\epsilon z_1(\rho,T,Z) \nonumber
\end{eqnarray}
Then, the off-diagonal terms in (\ref{equ01}) become,
\begin{eqnarray}
\label{gp04}
  g_{\rho T}(\rho,T,Z) &=& \epsilon \left(g^{1}_{rt}(\rho,T,Z)+g^0_{rr}(\rho)\left(\frac{\partial r_1}{\partial T}-\frac{\partial t_1}{\partial \rho}\right)\right)  \nonumber \\
  g_{\rho Z}(\rho,T,Z) &=& \epsilon \left(g^{1}_{rz}(\rho,T,Z)+g^0_{rr}(\rho)\frac{\partial r_1}{\partial Z}+g^0_{zz}(\rho)\frac{\partial z_1}{\partial \rho}\right)  \\
  g_{T Z}(\rho,T,Z) &=& \epsilon \left(g^{1}_{tz}(\rho,T,Z)-g^0_{rr}(\rho)\frac{\partial t_1}{\partial Z}+g^0_{zz}(\rho)\frac{\partial z_1}{\partial T}\right)  \nonumber
\end{eqnarray}
Then, setting all the left hand sides in (\ref{gp04}) to zero, we get a set of equations for $t_1$, $r_1$, and $z_1$ than can be straightforwardly solved for any choice of $g^{1}_{rt}$, $g^{1}_{rz}$ and $g^{1}_{tz}$. On this account, we take for the perturbed metric the form,
\begin{eqnarray}\label{gp05}
ds^2 & = & -e^{2\gamma-2 \psi}(1+\epsilon h_{tt}(r,t,z) ) dt^2 + e^{2\gamma-2 \psi}(1+\epsilon h_{rr}(r,t,z)) dr^2 \nonumber \\
 & & + e^{2 \psi}(1+\epsilon h_{zz}(r,t,z))dz^2 +e^{-2
\psi}r^2(1+\epsilon h_{\phi \phi}(r,t,z)) d\phi^2
\end{eqnarray}

Unfortunately, imposing the vanishing of the off-diagonal terms does not fix the gauge completely, and, therefore, (\ref{gp05}) is not unique.   It is easy to get the most general transformation of the form (\ref{gp02}) that preserves the diagonal form (\ref{gp05}). The corresponding coordinate transformation functions can be written in the form,
\begin{eqnarray}
\label{gp06}
  t_1(\rho,T,Z) &=& \frac{\partial W_1}{\partial T} +e^{2 \psi-\gamma}\frac{\partial W_3}{\partial T}   \nonumber \\
  r_1(\rho,T,Z) &=& \frac{\partial W_1}{\partial \rho} +\left(2\frac{d \psi}{d\rho} -\frac{d \gamma}{d\rho}\right) e^{2 \psi-\gamma} W_3
  +e^{4 \psi-2\gamma}\frac{\partial W_2}{\partial \rho} \\
  z_1(\rho,T,Z) &=&  e^{-2 \psi+\gamma}\frac{\partial W_3}{\partial Z}-\frac{\partial W_2}{\partial Z} \nonumber
\end{eqnarray}
where $W_1=W_1(\rho,T)$, $W_2=W_2(T,Z)$, and $W_3=W_3(\rho,Z)$ are arbitrary independent functions of the indicated arguments. If we consider the transformations generated by each one of these functions separately, starting with the background (order zero) metric, we have three cases \footnote{Notice that in all quantities that are already first order in $\epsilon$ the arguments may be taken as $(\rho,T,Z)$ or $(r,t,z)$ indistinctly, as this leads to differences that are of second order in $\epsilon$}  :

\subsection{Case $W_1(\rho,T) \neq 0$, $W_2(\rho,Z) = 0$ and $W_3(T,Z) = 0$.}

In this case we have to consider general perturbations that depend only on $(r,t)$, and, therefore, preserve the cylindrical symmetry of the unperturbed metric. The gauge independent part corresponds to cylindrical gravitational waves and will be excluded here, as we are interested only in perturbations that break that symmetry. From the point of view of the initial value problem, they correspond to initial data independent of $z$, and, therefore can be excluded by simply considering only initial data that is compactly supported in both $z$, and $r$.

\subsection{Case $W_1(\rho,T) = 0$, $W_2(\rho,Z) \neq 0$ and $W_3(T,Z) = 0$. ``Zero'' modes.}

This case corresponds to static perturbations (independent of $t$), i.e., ``zero'' modes. We first explore the general solution of the linearized Einstein equations by assuming for the perturbations the form,
\begin{eqnarray}
\label{scrz01}
  h_{tt}(r,t,z)  &=&   F_1(r,z)\nonumber \\
  h_{rr}(r,t,z)  &=&   F_2(r,z)\nonumber \\
  h_{zz}(r,t,z)  &=&   K_1(r,z)  \\
  h_{\phi \phi}(r,t,z)  &=&   K_2(r,z)\nonumber
\end{eqnarray}
Replacing in the linearized Einstein equations we find that,
\begin{equation}\label{scrz03}
 F_2=\frac{r}{(1+\kappa)^2} \frac{\partial F_1}{\partial r}+\frac{r}{(1+\kappa)^2} \frac{\partial K_2}{ \partial r}+\frac{\kappa(2+\kappa)}{(1+\kappa)^2}F_1+
 +\frac{(1+2\kappa)}{(1+\kappa)^2}K_2
\end{equation}
and,
\begin{equation}\label{scrz05}
 \frac{\partial K_1}{\partial r}=-\frac{\kappa^2}{(1+\kappa)^2} \frac{\partial K_2}{\partial r}-\frac{1}{(1+\kappa)^2} \frac{\partial F1}{\partial r}
 -\frac{1}{(1+\kappa)^2}r^{2\kappa^2+4\kappa+1}\left( \frac{\partial^2 F_1}{\partial z^2}+\frac{\partial^2 K_2}{\partial z^2}\right)
\end{equation}
There is only one further independent equation that can be written in the form,
\begin{equation}\label{scrz07}
    \frac{\partial^2 Q}{\partial r^2}+\frac{1}{r}\frac{\partial Q}{ \partial r} + r^{2\kappa(2+\kappa)} \frac{\partial^2 Q}{\partial z^2} =0
\end{equation}
where,
\begin{equation}\label{scrz09}
    Q(r,z)= F_1(r,z)-\kappa K_2(r,z)
\end{equation}
Thus, the general $t$ independent solution is determined by a solution of (\ref{scrz07}) plus an arbitrary function of $(r,z)$. To see the relation of this result to the gauge ambiguities we notice that under a coordinate transformation where only $W_2(\rho,Z) \neq 0$, we have that the relevant functions $h_{ab}$ transform as,
\begin{eqnarray}
\label{scrz11}
  F_1(r,z)  & \to &  F_1(r,z)+ 2 \kappa(1+\kappa)\; r^{-2\kappa^2-4\kappa-1} \frac{\partial W_2}{\partial r}     \nonumber \\
  F_2(r,z)  & \to &   F_2(r,z) - 2\kappa(3+\kappa)\;r^{2\kappa^2-4\kappa-1}\frac{\partial W_2}{\partial r}
  +2 r^{-2\kappa(2+\kappa)} \frac{\partial^2 W_2}{\partial r^2}
  \nonumber \\
  K_1(r,z)  & \to &  K_1(r,z) -2 \kappa \; r^{-2\kappa^2-4\kappa-1}\frac{\partial W_2}{\partial r} -2 \frac{\partial^2 W_2}{\partial z^2}   \\
  K_2(r,z)  & \to &  K_2(r,z)+2 (1+\kappa)\; r^{-2\kappa^2-4\kappa-1} \frac{\partial W_2}{\partial r}  \nonumber
\end{eqnarray}
The transformation law for $F_1$ and $K_2$ implies that the function $Q(r,z)$ is gauge invariant, and, therefore, it represents a gauge non trivial perturbation. To analyze this in more detail, we notice that if we write,
\begin{equation}\label{scrz14}
    Q(r,z)=\int{e^{-ikz}\widetilde{Q}(k,r)} dk\;,
\end{equation}
then, $\widetilde{Q}(k,r)$ satisfies the equation,
\begin{equation}\label{scrz16}
    \frac{d^2 \widetilde{Q}}{d r^2}+\frac{1}{r}\frac{d \widetilde{Q}}{ r} -k^2 r^{2\kappa(2+\kappa)} \widetilde{Q} =0 \; ,
\end{equation}
whose general solution is,
\begin{equation}\label{scrz18}
          \widetilde{Q} (k,r) = C_1 I_0 \left(k r^{(1+\kappa)^2}/(1+\kappa)\right)+C_2 K_0 \left(k r^{(1+\kappa)^2}/(1+\kappa)\right)
\end{equation}
where $C_1$, and $C_2$, are constants, and $I_0(x)$, and $K_0(x)$ are modified Bessel functions of the first and second kind, respectively. But, given the properties of these functions, this result implies that there are no gauge invariant, time independent perturbations that are finite in both limits $r \to 0$ and $r \to \infty$. We may again exclude them by restricting to initial data of compact support in both $(r,z)$.

\subsection{Case $W_1(\rho,T) = 0$, $W_2(\rho,Z) = 0$ and $W_3(T,Z) \neq 0$.}

This type of transformation is relevant in the general problem where the perturbations effectively depend on $(r,t,z)$. The result of applying it on the perturbed metric  (\ref{gp05}) is,
\begin{eqnarray}
\label{sc03}
  h_{tt} & \to & h_{tt}-2\kappa^2(1+\kappa)(2+\kappa)r^{-2-2\kappa-\kappa^2}W_3(t,z)+2r^{-\kappa(2+\kappa)} \frac{\partial^2 W_3}{\partial t^2}
  \nonumber \\
  h_{rr} & \to & h_{rr}+2\kappa (1+\kappa)(2+\kappa)r^{-2-2\kappa-\kappa^2}W_3(t,z)  \\
  h_{zz} & \to & h_{zz}+2\kappa^2 (2+\kappa)r^{-2-2\kappa-\kappa^2}W_3(t,z)+2r^{\kappa(2+\kappa)} \frac{\partial^2 W_3}{\partial z^2}
  \nonumber \\
   h_{\phi \phi} & \to & h_{\phi \phi}-2\kappa (1+\kappa)(2+\kappa)r^{-2-2\kappa-\kappa^2}W_3(t,z)  \nonumber
\end{eqnarray}

Since this preserves the general form (\ref{gp05}), any result obtained by solving Einstein's equations for the $h_{ab}$ will be subject to a gauge ambiguity. Nevertheless, we may extract from (\ref{sc03}) the following gauge invariant quantities,
\begin{eqnarray}
\label{sc30}
  \widetilde{G}_1(r,t,z) & = &  h_{zz} + \frac{\kappa}{\kappa+1}h_{\phi\phi} +\frac{r^{2(1+\kappa)^2}}{\kappa(1+\kappa)(2+\kappa)} \frac{\partial^2 h_{\phi\phi}}{\partial z^2}
  \nonumber \\
  \widetilde{G}_2(r,t,z) & = &  \frac{1}{1+\kappa} h_{tt} +h_{zz}+\frac{r^{2(\kappa+1)^2}}{\kappa(1+\kappa)(2+\kappa)} \frac{\partial^2 h_{\phi\phi}}{\partial z^2} +\frac{r^2}{\kappa(1+\kappa)^2(2+\kappa)}\frac{\partial^2 h_{\phi\phi}}{\partial t^2}
   \\
 \widetilde{G}_3(r,t,z) & = &    h_{tt} -\kappa h_{\phi\phi}  +\frac{r^2}{\kappa(1+\kappa)(2+\kappa)}\frac{\partial^2 h_{\phi\phi}}{\partial t^2} \nonumber
\end{eqnarray}
Actually, these three functions are not independent, and we have,
\begin{equation}\label{sc32}
     \widetilde{G}_1 - \widetilde{G}_2 -\frac{1}{1+\kappa} \widetilde{G}_3=0
\end{equation}

Its implications in deriving relevant physical conclusions will be considered in the following Sections.

\section{Perturbations along the symmetry axis II. The linearized equations.}

As discussed in the previous Sections, we may restrict the perturbations to diagonal terms. Furthermore, since the unperturbed metric depends only on $r$, and the equations are linear, it will be appropriate to assume that the dependence on $t$ and $z$ is only through a factor $\exp(i(\Omega t - k z))$, with the most general solution a linear combination of the resulting solutions. Thus we take for the perturbed metric the form,
\begin{eqnarray}\label{equ02}
ds^2 & = & -e^{2\gamma-2 \psi}(1+\epsilon e^{i(\Omega t - k z)}F_1(r)) dt^2 + e^{2\gamma-2 \psi}(1+\epsilon e^{i(\Omega t - k z)}F_2(r)) dr^2 \nonumber \\
 & & + e^{2 \psi}(1+\epsilon e^{i(\Omega t - k z)} K_1(r))dz^2 +e^{-2
\psi}r^2(1+\epsilon e^{i(\Omega t - k z)} K_2(r)) d\phi^2
\end{eqnarray}
where, as usual, it is understood that one should take only the real part of the full complex expression.

We next impose that (\ref{equ02}) satisfies the vacuum Einstein equations to first order in $\epsilon$. This leads in general to a set of coupled second order ordinary differential equations for the functions $F_i$ and $K_i$, but it is easy to show that we must have,
\begin{equation}\label{equ09}
    F_2(r)=-K_2(r)
\end{equation}
and one can reduce the rest of the system to the following set of coupled first order equations,
\begin{eqnarray}
\label{equ10}
  \frac{dF_1}{dr} &=& \frac{\left(r^2\Omega^2+\kappa(2+\kappa)(1+\kappa)^2\right)}{2 r (1+\kappa)}K_1
                     +\frac{\left(k^2 r^{2(1+\kappa)^2}-\kappa(2+\kappa)(2\kappa+1)\right)}{2 r (1+\kappa)}F_1 \nonumber \\
                   & &  +\frac{\left(k^2 r^{2(1+\kappa)^2}-4\kappa(1+\kappa)^2 -\Omega^2 r^2-\kappa^4\right)}{2 r (1+\kappa)}K_2
 \nonumber \\
  \frac{dK_1}{dr} &=& -\frac{\left(r^2\Omega^2+\kappa(2+\kappa)(\kappa^2-1)\right)}{2 r (1+\kappa)}K_1
                     +\frac{\left(k^2 r^{2(1+\kappa)^2}+\kappa(2+\kappa)\right)}{2 r (1+\kappa)}F_1 \nonumber \\
                   & &  +\frac{\left(k^2 r^{2(1+\kappa)^2}-\Omega^2 r^2+4\kappa+4\kappa^2  -\kappa^4\right)}{2 r (1+\kappa)}K_2
 \\
 \frac{dK_2}{dr} &=& \frac{\left(r^2\Omega^2+\kappa(2+\kappa)(1+\kappa)^2\right)}{2 r (1+\kappa)}K_1
                     -\frac{\left(k^2 r^{2(1+\kappa)^2}+\kappa(2+\kappa)\right)}{2 r (1+\kappa)}F_1 \nonumber \\
                   & &  -\frac{\left(k^2 r^{2(1+\kappa)^2}-\Omega^2 r^2-4+8\kappa+2\kappa^2 -2\kappa^3 -\kappa^4\right)}{2 r (1+\kappa)}K_2
 \nonumber
\end{eqnarray}

What we have in mind here is that, with appropriate boundary conditions, the set of solutions of (\ref{equ10}), together with the $e^{-ikz}$ factors, will provide a basis for an expansion of arbitrary functions of $(r,z)$, leading to the formal solution of the initial value problem for the perturbations. On this account we may look at (\ref{equ10}) as a boundary value problem that determines the allowed values of $\Omega$, such that the boundary conditions are satisfied. As we show in the next Section, it turns out to be convenient for this purpose to change (\ref{equ10}) to an equation for a single function, that satisfies a third order ordinary differential equation. An issue that will also require further discussion is that of the gauge dependence of the solutions.

\section{The general solution}

Considering again the system (\ref{equ10}), we notice that, if we are interested in the initial value problem for the perturbations, in particular if we are considering the possibility of unstable modes, then, the question is: given $\kappa$ and $k$, are there solutions with acceptable values of $\Omega$? That is, are there values of $\Omega$ such that the solutions satisfy appropriate boundary conditions? This is a standard boundary value problem, but it is not easy to handle if given in the form (\ref{equ10}). A simpler problem is obtained if introduce two new functions, $G_1(r)$ and $G_2(r)$, such that,
\begin{equation}\label{gs02}
    K_1(r)=\left(G_1(r)+G_2(r)\right)/2\;\;\;;\;\;\;K_2(r)=\left(G_1(r)-G_2(r)\right)/2
\end{equation}
Then, using (\ref{equ10}) repeatedly, we find,
\begin{eqnarray}
\label{gs04}
  G_2(r) &=& -\frac{\kappa^2+\kappa-1}{1+\kappa} G_1(r)+\frac{r}{1+\kappa} \frac{dG_1}{dr} \nonumber \\
  F_1(r) &=& \frac{r^2}{k^2 r^{2(1+\kappa)^2}+2\kappa+\kappa^2} \frac{d^2G_1}{dr^2} + \frac{r\left(6+8\kappa+\kappa^2+ k^2 r^{2(1+\kappa)^2}\right)}{2(1+\kappa)\left(k^2 r^{2(1+\kappa)^2}+2\kappa+\kappa^2\right)} \\
   & &  -\frac{4\kappa+10\kappa^2+6\kappa^3+\kappa^4 -2 r^2(1+\kappa)\Omega^2+\kappa(2+\kappa)k^2 r^{2(1+\kappa)^2}}{2(1+\kappa)\left(k^2 r^{2(1+\kappa)^2}+2\kappa+\kappa^2\right)} G_1(r)
\end{eqnarray}
while for $G_1$ we find the equation,
\begin{eqnarray}
\label{gs06}
  \frac{d^3 G_1}{dr^3} &=& \frac{k^2A(k^2A+6\kappa^2+12\kappa+3)-(k^2 A+2\kappa+\kappa^2)r^2 \Omega^2 -\kappa(\kappa^3+4\kappa+6+7\kappa)}
   {r^2(k^2A+2\kappa+\kappa^2)} \frac{dG_1}{dr} \nonumber \\
   & &  +\frac{(\kappa+3)(\kappa-1)k^2 A -\kappa^4-4\kappa^3-9\kappa^2-10\kappa}{r(k^2A+2\kappa+\kappa^2)} \frac{d^2G_1}{dr^2}\\
   & & + \frac{\kappa(2+\kappa)((k^2A -\kappa^2-2\kappa-2)r^2\Omega^2-k^2A(k^2A+2+6\kappa+3\kappa^2))}{r^3(k^2A+2\kappa+\kappa^2)} G_1
   \nonumber
\end{eqnarray}
where,
\begin{equation}\label{gs07}
    A=r^{2(1+\kappa)^2}
\end{equation}

It may appear rather unexpected that (\ref{gs06}) has an exact solution given by,
\begin{equation}\label{gs08}
    G_1(r)=\frac{\kappa(\kappa+2)+k^2r^{2(1+\kappa)^2}}{r^{2+2\kappa+\kappa^2}}
\end{equation}
Notice that (\ref{gs08}) is independent of $\Omega$. One can show, however, following the discusion in Section III.C, that this solution is a consequence of the gauge ambiguity intrinsic in the system (\ref{equ10}). To see how this happens we recall that $G_1$ is defined up to constant and write,
\begin{equation}\label{gs08a}
    G_1(r)=w(\Omega,k)\frac{\kappa(\kappa+2)+k^2r^{2(1+\kappa)^2}}{r^{2+2\kappa+\kappa^2}}
\end{equation}
where $w(\Omega,k)$ is an arbitrary function of $\Omega$, and $k$. Replacing in (\ref{gs02}), and (\ref{gs04}), we find, (recall that $F_2=-K_2$),
\begin{eqnarray}
\label{gs10}
  F_1(r) &=& \kappa^2 (1+\kappa)(2+\kappa) r^{-2-2\kappa-\kappa^2} w(\Omega,k) +\Omega^2 r^{-\kappa(2+\kappa)} w(\Omega,k) \nonumber \\
  K_1(r) &=&  - \kappa^2 (2+\kappa) r^{-2-2\kappa-\kappa^2} w(\Omega,k) + k^2 r^{\kappa(2+\kappa)} w(\Omega,k) \\
  K_2(r) &=& \kappa(1+\kappa)(2+\kappa) r^{-2-2\kappa-\kappa^2} w(\Omega,k) \nonumber
\end{eqnarray}
Upon multiplication by $e^{i(\Omega t - k z)}$ and integration on $\Omega$ and $k$, we recover the full $(r,t,z)$ dependence, and we get,
\begin{eqnarray}
\label{gs12}
  F_1(r,t,z) &=& \kappa^2 (1+\kappa)(2+\kappa) r^{-2-2\kappa-\kappa^2} W(t,z) - r^{-\kappa(2+\kappa)} \frac{\partial^2 W}{\partial t^2} \nonumber \\
  K_1(r,t,z) &=&  - \kappa^2 (2+\kappa) r^{-2-2\kappa-\kappa^2} W(t,z) - r^{\kappa(2+\kappa)} \frac{\partial^2 W}{\partial z^2} \\
  K_2(r,t,z) &=& \kappa(1+\kappa)(2+\kappa) r^{-2-2\kappa-\kappa^2} W(t,z) \nonumber
\end{eqnarray}
where,
\begin{equation}\label{gs14}
    W(t,z)=\int\int e^{i(\Omega t - k z)} w(\Omega,k) d\Omega\; dk
\end{equation}
But a comparison with (\ref{sc03}) immediately shows that this solution is pure gauge and should be discarded. Nevertheless, we may use it to look for solutions of (\ref{gs06}) of the form,
\begin{equation}\label{gs10a}
    G_1(r)=\frac{\kappa(\kappa+2)+k^2r^{2(1+\kappa)^2}}{r^{2+2\kappa+\kappa^2}} H_1(r)
\end{equation}
Replacing in (\ref{gs06}), we find that $H_1(r)$ satisfies the equation,
\begin{eqnarray}
\label{gs12a}
  & & -\frac{d^3 H_1}{dr^3} -\frac{(3+4\kappa+2\kappa^2)k^2A-\kappa(9\kappa+2+2\kappa^3+8\kappa^2)}{r(k^2A+\kappa^2+2\kappa)}\frac{d^2H_1}{dr^2} \nonumber \\
  & &  + \left[ \frac{k^4A^2(k^2A+8\kappa-4\kappa^3+3-\kappa^4)-\kappa(\kappa+2)(5\kappa(2+\kappa)(3+2\kappa^2+4\kappa)+6)k^2A } {r^2(k^2A+\kappa^2+2\kappa)^2}\right. \nonumber \\
  & &  -\left. \frac{\kappa^2(2+\kappa)^2(1+\kappa)^4} {r^2(k^2A+\kappa^2+2\kappa)^2}\right]\frac{dH_1}{dr} \nonumber \\
   && =\Omega^2 \frac{dH_1}{dr}
\end{eqnarray}
To simplify further the treatment, we introduce a new function $H_2(r)$ such that,
\begin{equation}\label{gs14a}
   \frac{dH_1(r)}{dr}= {\cal{K}}(r) H_2(r)
\end{equation}
where,
\begin{equation}\label{gs16}
    {\cal{K}}(r)= \frac{\sqrt{r} r^{2\kappa+\kappa^2}}{\kappa(\kappa+2) + k^2 r^{2(1+\kappa)^2}}
\end{equation}
Replacing in (\ref{gs12}) we get an equation for $H_2(r)$ of the form,
\begin{equation}\label{gs17}
    -\frac{d^2H_2}{dr^2}+V_2 H_2(r) = \Omega^2 H_2(r)
\end{equation}
where $V_2$ is a function of $r$, $\kappa$, and $k$. It will be convenient to introduce a further change of variable and a new function given by,
\begin{eqnarray}
\label{gs18}
  x(r) &=& |k|^{\frac{1}{(1+\kappa)^2}} r  \nonumber \\
   H_2(r) &=& H_3(x(r))
\end{eqnarray}
and we find that $H_3(x)$ satisfies the equation,
\begin{equation}\label{gs20}
    -\frac{d^2H_3}{dx^2}+V_3(x) H_3(x) = \lambda H_3(x)
\end{equation}
where,
\begin{equation}\label{gs22}
    \lambda= \frac{\Omega^2}{|k|^{\frac{2}{(1+\kappa)^2}}}
\end{equation}
and,
\begin{eqnarray}
\label{gs24}
   V_3(x) &=&  \left[-{\kappa}^{2} \left( 2+\kappa \right) ^{2}+4\,{x}^{6\, \left( 1+\kappa \right) ^{2}}+ \left( 48\,\kappa+15+24\,{
\kappa}^{2} \right) {x}^{4\, \left( 1+\kappa \right) ^{2}} \right. \nonumber \\
  &  & \left.-2\,\kappa\,
 \left( 2+\kappa \right)  \left( 16\,{\kappa}^{4}+64\,{\kappa}^{3}+86
\,{\kappa}^{2}+44\,\kappa+9 \right) {x}^{2\, \left( 1+\kappa \right) ^
{2}}
 \right]\\
 & & \times\left[4 x^2\left(x^{2(1+\kappa)^2}+\kappa(2+\kappa)\right)^2\right]^{-1} \nonumber
\end{eqnarray}

We cannot fail to notice that (\ref{gs22}) has the form of a ``dispersion relation'' ($\Omega = |k|^{\frac{2}{(1+\kappa)^2}}\sqrt{\lambda} $), that changes with $\lambda$. We remark, however, that $k$ is not the modulus of the ``wave vector''. In principle, one can understand (\ref{gs22}) as resulting from the interference between a traveling wave part in the $z$ direction and a standing wave in the radial direction that characterize the modes described by $H_3$ \\

Equation (\ref{gs20}) has the typical form of an eigenvalue - eigenfunction equation. We first need to establish the general behaviour of its solutions both for $x \to 0$, and $x \to \infty$. Since (\ref{gs20}) is a second order ordinary differential equation, it has two independent solutions. Let us consider first $r \to 0$. It is not difficult to show that near $r=0$, the general solution of (\ref{gs20}) admits an asymptotic expansion of the form,
\begin{equation}\label{gs26}
    H_3(x)=\sqrt{x} \sum_{j=0} x^{2j(1+\kappa)^2}\left[ \sum_{i=0}  a^{(2j)}_{2i} x^{2i} +\ln(x)\sum_{i=0}  b^{(2j)}_{2i} x^{2i}\right]
\end{equation}
or, in more detail, up to leading order terms in $x$,
\begin{eqnarray}
\label{gs28}
  H_3(x) &=& \sqrt{x} \left[a^{(0)}_0 +\frac{\lambda}{4} \left(b^{(0)}_0-a^{(0)}_0 \right) x^2  +...
                     +\ln(x) \left(b^{(0)}_0 -\frac{\lambda}{4} b^{(0)}_0x^2 +...\right)\right. \nonumber \\
   & &  \left.+x^{2(1+\kappa)^2} \left[a^{(2)}_0 + a^{(2)}_2 x^2 +  ...
                     +\ln(x) \left(b^{(2)}_0 + b^{(2)}_2x^2 +  ...\right)\right]  +\dots \right]
\end{eqnarray}
where the dots indicate higher terms. All the $b^{(2j)}_{2i}$ coefficients are proportional to $b^{(0)}_0$, and the $a^{(2j)}_{2i}$ are linear homogeneous in $a^{(0)}_0$, and $b^{(0)}_0$. This implies that we have two sets of independent solutions that can be selected by choosing either $b^{(0)}_0=0$ or $a^{(0)}_0=0$. In the first case, ($b^{(0)}_0=0$), the solutions behave as $\sqrt{x}+{\cal{O}}(x^{5/2})$ as $x \to 0$, and in the second, ($a^{(0)}_0=0$), the solutions behave as $\ln(x)\left(\sqrt{x}+{\cal{O}}(x^{5/2}) \right)$ as $x \to 0$.

In accordance with (\ref{gs10}), (\ref{gs14}), (\ref{gs16}), and (\ref{gs18}), we have that the general solution of (\ref{gs06}) may be written in the form,
\begin{equation}\label{gs40}
    G_1(r) = \frac{k^2 r^{2(1+\kappa)^2}+\kappa(\kappa+2)}{r^{2+2\kappa+\kappa^2}} \left[ \int_0^r{\frac{\sqrt{r_1}r_1^{2\kappa+\kappa^2}}{k^2 r_1^{2(1+\kappa)^2}+\kappa(\kappa+2)} H_3(|k|^{1/(1+\kappa)^2}r_1) dr_1}+ C_{G_1}\right]
\end{equation}
where $H_3(x)$ is the general solution of (\ref{gs20}), and $C_{G_1}$ is a constant. With this result and the expansion (\ref{gs28}), we conclude that besides the solution (\ref{gs08}), $G_1(r)$ has two other linearly independent solutions, which, near $r=0$, and up to terms of order $r^2$, may be written in the form,
\begin{equation}\label{gs42}
    G_1(r) \sim b_0\left(1 + b_2 r^2+\dots \right)+c_0\left[1+c_2r^2+\dots+\ln(r)\left(d_0+d_2 r^2+\dots\right)\right]
\end{equation}
where $b_2$, $c_2$, $d_0$ and $d_2$ depend on $\Omega$ and $\kappa$, and are finite. Thus we get one solution that is finite for $r=0$ by setting $b_0\neq 0\;,\;c_0=0$, and an independent solution that diverges as $\ln(r)$ by setting $b_0 = 0\;,\;c_0\neq 0$.

The behaviour of $H_3(x)$ for large $x$ is more difficult to establish, and we have not been able to obtain asymptotic expansions similar to (\ref{gs26}) or (\ref{gs28}). We nevertheless notice that for large $x$ we have,
\begin{equation}\label{gs50}
    V_3(x) \sim x^{2\kappa(2+\kappa)}
\end{equation}
and we may approximate (\ref{gs20}) by,
\begin{equation}\label{gs52}
   -\frac{d^2 H_3}{dx^2} +x^{2\kappa(2+\kappa)} H_3(x) = 0
\end{equation}
The general solution of this equation is,
\begin{equation}\label{gs54}
    H_3(x)=C_1 \sqrt{x} I_{(2(1+\kappa)^2)}\left(\frac{x^{(1+\kappa)^2}}{(1+\kappa)^2}\right)+C_2 \sqrt{x} K_{(2(1+\kappa)^2)}\left(\frac{x^{(1+\kappa)^2}}{(1+\kappa)^2}\right)
\end{equation}
where $I_{\alpha}(x)$ and $K_{\alpha}(x)$ are modified Bessel functions. Actually (\ref{gs54}) is consistent only if we keep the leading terms. Therefore, for $x \to \infty$ we have two linearly independent solutions for $H_3(x)$, whose leading terms are,
\begin{equation}\label{gs56}
    H_3(x)=C_1 \sqrt{x} \exp\left({\frac{x^{(1+\kappa)^2}}{(1+\kappa)^2}}\right)+C_2 \sqrt{x} \exp\left({-\frac{x^{(1+\kappa)^2}}{(1+\kappa)^2}}\right)
\end{equation}

We need to establish now what boundary conditions should be considered acceptable. In principle, we expect a perturbation to be finite everywhere and that it should be possible to make it arbitrarily smaller than the background. On this account, considering (\ref{gs40}), the only acceptable solutions for $r \to 0$ are those where we take the solution for $H_3$ that vanishes as $\sqrt{x}$, and $C_{G_1}=0$.

To analyze the behaviour for large $r$ we write (\ref{gs40}), (with $C_{G_1}=0$), in the form,
\begin{eqnarray}\label{gs40a}
    G_1(r) & = & -\frac{r^{2(1+\kappa)^2}+\kappa(\kappa+2)}{r^{2+2\kappa+\kappa^2}} \int_r^{\infty}{\frac{\sqrt{r_1}r_1^{2\kappa+\kappa^2}}{r_1^{2(1+\kappa)^2}+\kappa(\kappa+2)} H_3(|k|^{1/(1+\kappa)^2}r_1) dr_1} \nonumber \\
    & & + \frac{r^{2(1+\kappa)^2}+\kappa(\kappa+2)}{r^{2+2\kappa+\kappa^2}} I_3
\end{eqnarray}
where,
\begin{equation}\label{gs40b}
    I_3 =  \int_0^{\infty}{\frac{\sqrt{r_1}r_1^{2\kappa+\kappa^2}}{r_1^{2(1+\kappa)^2}+\kappa(\kappa+2)} H_3(|k|^{1/(1+\kappa)^2}r_1) dr_1}
\end{equation}

Therefore, if we impose the condition that $H_3(x)$ should vanish for $x \to \infty$, the integral $I_3$ will be finite, and as $x \to \infty$, the first term in (\ref{gs40a}) vanishes, and the remaining term, in accordance with our previous discussion, is pure gauge. Thus, we conclude that the appropriate boundary conditions for $H_3(x)$ are,
\begin{eqnarray}
\label{gs40c}
  H_3(x) &\sim & \sqrt{x} \;\;\; , \;\;\; {\mbox{for  }} \; x \to 0 \nonumber \\
  H_3(x) & \to &  0 \;\;\;\;\;, \;\;\;\; {\mbox{for  }} \; x \to \infty
\end{eqnarray}

The consequences of these boundary conditions for $H_3$ on its spectrum, and therefore on the behaviour of the perturbations will be analyzed in the next Section.

\section{The spectrum of $\Omega$ and unstable solutions}

As indicated in the previous Section, we need to impose boundary conditions on $G_1(r)$ to restrict to geometrically acceptable perturbations. This, in turn, imposes boundary conditions on $H_3(x)$. These conditions are that $H_3$ should go to zero for $x \to \infty$ and that it should vanish as $\sqrt{x}$ for $x \to 0$. We recall that $H_3$ is a solution of (\ref{gs20}). With the stated boundary conditions the operator on the left of (\ref{gs20}) is self-adjoint, and, since $V_3 \to +\infty$ for $x \to +\infty$, its spectrum (i.e. the set of allowed values of $\lambda$) is fully discrete. We may classify the corresponding eigenfunctions by the number of nodes $j$, that is, the zeros of $H_3$ for $x \neq 0$, and indicate them as $H_3^{(j)}(x)$. We then have,
\begin{equation}\label{ts02}
    -\frac{d^2H_3^{(j)}}{dx^2}+V_3(x) H_3^{(j)}(x) = \lambda_j H_3^{(j)}(x)
\end{equation}
where $\lambda_j$ is the corresponding eigenvalue.

Since the spectrum is discrete, the eigenfunctions $H_3^{(j)}$ are real and can be normalized so that they satisfy the relations,
\begin{equation}\label{ts04}
    \int_0^{\infty}{H_3^{(j)}(x)H_3^{(\ell)}(x) dx} = \delta_{j \ell}
\end{equation}
With this normalization the set $\left\{H_3^{(j)}(x)\;;\;j=0,1,2\dots\right\}$ provides a complete orthonormal basis for the expansion of functions of $x$ in $0 \leq x < \infty$.

We recall that,
\begin{equation}\label{ts06}
    \Omega^2= |k|^{\frac{2}{(1+\kappa)^2}} \lambda
\end{equation}
Therefore, a negative $\lambda$ corresponds to a pure imaginary value of $\Omega$, and this, in turn, to unstable perturbations that diverge exponentially in time. Because of the rather involved dependence of $V_3$ on $x$, it is not easy to determine analytically the details of the spectrum of $\lambda$. We may, nevertheless, resort to numerical methods to try to determine at least the lowest values of $\lambda$, remembering that they still depend on $\kappa$.

As a simple numerical approach we imposed on $H_3(x)$ the boundary condition (\ref{gs28}), with $b^{(0)}_0=0$, and, using a ``shooting'' method, we solved the boundary value problem for (\ref{ts02}) for the lowest eigenvalues, finding approximate values for $\lambda$, and corresponding functional forms for $H_3(x)$, for several choices of $\kappa$. The results for $\lambda_0$ and $\lambda_1$ are shown in Table I. \\

\begin{table}[!ht]
\begin{tabular}{|c|c|c|}
   \hline
   $\kappa$ & $\lambda_0$ & $\lambda_1$\\
   \hline
   0.001 & -1.00446265 &1.01856808 \\
   0.005 & -1.0062283 &1.0784925 \\
   0.01&  -1.01146 &1.14635835 \\
 0.05&  -1.06132 &1.653817\\
 0.1 & -1.137215 & 2.303944 \\
 0.5 & -2.38340712 & 6.55762 \\
 1.0 & -6.928205 &6.928165 \\
  2.0 & -44.50382 & 5.56061 \\
  2.5 & -91.390175 &5.331185 \\
   \hline
\end{tabular}
\caption{ Approximate values for $\lambda_0$ and $\lambda_1$ for several choices of $\kappa$.}
\end{table}

Some examples of the functional forms of $H_3(x)$ are given in Figure 1, and the potential $V_3(x)$ is plotted in Figure 2 for several values of $\kappa$. It is apparent from these plots that $V_3(x)$ is strongly dependent on $\kappa$, and this, in turn is reflected in the dependence of $\lambda_j$ on $\kappa$.\\

\begin{figure}
\centerline{\includegraphics[height=12cm,angle=-90]{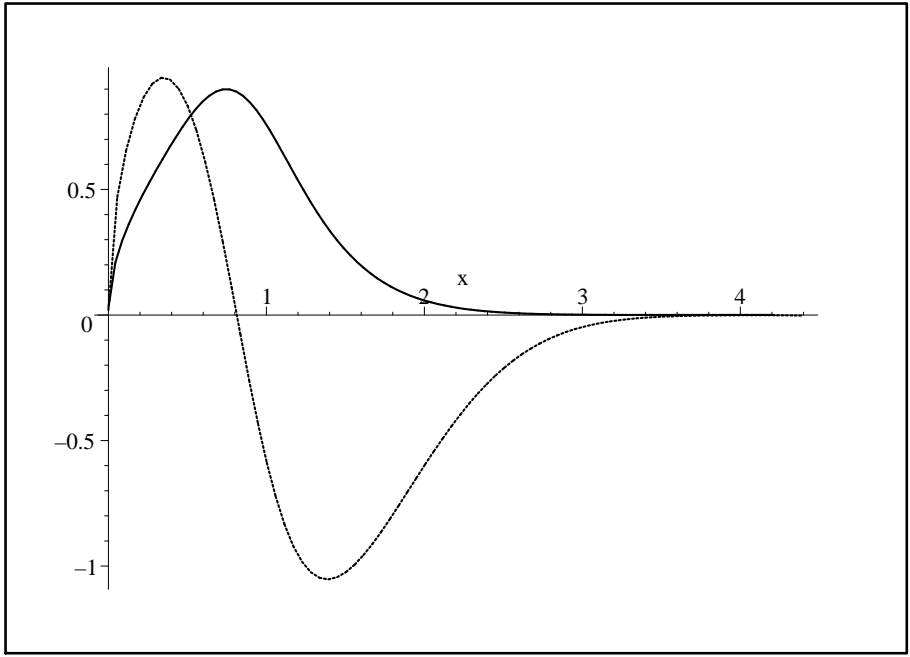}}
\caption{$H_3^{(0)}(x)$ as a function of $x$, for $\kappa=0.5$ and $\lambda_0=-2.38340712...$, (solid curve, no nodes), and $H_3^{(1)}(x)$, also for $\kappa=0.5$ and $\lambda_1=6.55762...$, (dotted curve, one node). The functions are not normalized.}
\end{figure}

\begin{figure}
\centerline{\includegraphics[height=12cm,angle=-90]{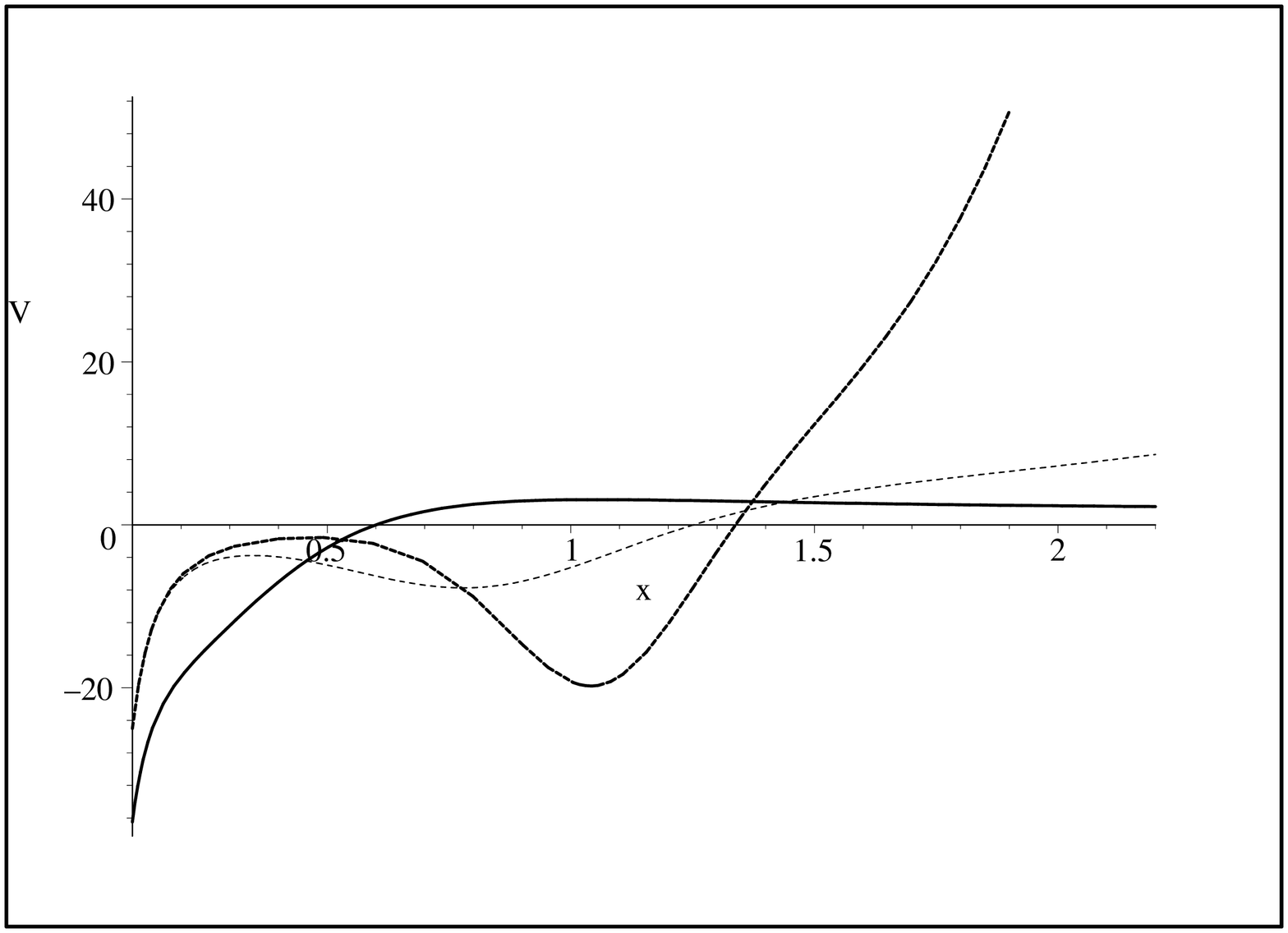}}
\caption{$V_3(x)$ as a function of $x$, for $\kappa=0.1$ (solid curve),  for $\kappa=0.5$ (dotted curve), and for $\kappa=1.0$ (dashed curve).}
\end{figure}

The most important feature of these results is that, in {\em all} cases considered, $\lambda_0 < 0$, and, therefore, we have shown that there is one {\em unstable } set of solution for the equations of motion for the perturbations for every value of $\kappa >0$ in the range of $\kappa$ considered. Although we do not have analytic results, we may prove that this result must hold for any value of $\kappa$ as follows. Suppose that there is a solution of $H_3^{(j)}(x)$ of (\ref{ts02}) with $\lambda_j =0 $. This implies that $\Omega_j =0$ for all $k$, and, therefore, replacing in (\ref{gs40}) we would obtain a gauge non trivial, time independent solution, that is finite in both limits $r \to 0$ and $r \to \infty$. But we have already shown that there are no such solutions, and, therefore, (\ref{ts02}) cannot have solutions with $\lambda=0$, satisfying the boundary conditions (\ref{gs40c}). But all $\lambda_j$, and, in particular, $\lambda_0$ and $\lambda_1$, are continuous functions of $\kappa$, and since they cannot vanish for any $\kappa$, for all $\kappa$ we must have $\lambda_0 < 0$, and $\lambda_1 >0$, and we conclude that for all $\kappa$ we have one, and only one negative eigenvalue $\lambda$.

\section{The initial value problem}

In this Section we consider the initial value problem for the perturbed system, but, before we carry out this analysis we need to see in what way the general problem of evolution is solved once we determine the allowed functions $H_3$ and the corresponding spectrum of values of $\lambda$. Let us consider again the functions $F_1(r)$, $F_2(r)$, $K_1(r)$, and $K_2(r)$. We have found that for every choice of $k$, after imposing appropriate boundary conditions, we have an infinite set of solutions, each one characterized by a function $H_3^{(j)}(x)$ and a value of $\lambda_j$. The allowed values of $\Omega$ follow from (\ref{ts06}).
\begin{equation}\label{ts06a}
   \Omega_j(k)= \pm |k|^{\frac{1}{(1+\kappa)^2}}\sqrt{\lambda_j}
\end{equation}
In what follows we shall call $ \Omega_j(k)$ the quantity corresponding to taking the plus sign in (\ref{ts06a}), and add a minus sign explicitly when necessary.

If we identify the corresponding metric coefficients as $F_1^{(j)}(k,r)$, $F_2^{(j)}(k,r)$,  $K_1^{(j)}(k,r)$, and $K_2^{(j)}(k,r)$, where we remark that these functions depend only on $k^2$, we may construct more general solutions by taking linear combinations. For example, we would have,
\begin{eqnarray}\label{ivp02}
    h_{zz}(r,t,z) & = & \sum_j{\int{\left(C^{(+)}_j(k) e^{i(\Omega_j(k)t-kz)}+C^{(-)}_j(k) e^{i(-\Omega_j(k)t-kz)} \right) K_1^{(j)}(k,r) dk}} \nonumber \\
    h_{\phi \phi}(r,t,z) & = & \sum_j{\int{\left(C^{(+)}_j(k) e^{i(\Omega_j(k)t-kz)}+C^{(-)}_j(k) e^{i(-\Omega_j(k)t-kz)} \right) K_2^{(j)}(k,r) dk}}
\end{eqnarray}
and similar expressions, with the same coefficients  $C^{(\pm)}_j(k)$, for $h_{tt}(r,t,z)$ and $h_{rr}(r,t,z)$. We may invert partially (\ref{ivp02}), multiplying by $e^{ikz}$ and integrating over $z$. Furthermore, we may set $t=0$ in both the functions and their $t$ derivatives, and finally combine them as follows,
\begin{eqnarray}\label{ivp04}
   & & \int{e^{ikz} \left(h_{zz}(r,0,z) + h_{\phi\phi}(r,0,z)\right) dz} \nonumber \\
    & & = \;\;\;\;\;\;\; 2 \pi\sum_j{\left(C^{(+)}_j(k)+C^{(-)}_j(k) \right) \left(K_1^{(j)}(k,r) + K_2^{(j)}(k,r)\right)}
\end{eqnarray}
and,
\begin{eqnarray}\label{ivp06}
   & & \int{e^{ikz} \left(\left.\frac{\partial h_{zz}}{\partial t}\right|_{t=0} + \left.\frac{\partial h_{\phi\phi}}{\partial t}\right|_{t=0}\right) dz} \nonumber \\
    & & = \;\;\;\;\;\;\; 2 \pi i\sum_j{\Omega_j(k)\left(C^{(+)}_j(k)-C^{(-)}_j(k) \right) \left(K_1^{(j)}(k,r) + K_2^{(j)}(k,r)\right)}
\end{eqnarray}
Then, if we define,
\begin{eqnarray}\label{ivp08}
   {\cal{I}}_1(k,r) & = & \int{e^{ikz} \left(h_{zz}(r,0,z) + h_{\phi\phi}(r,0,z)\right) dz} \nonumber \\
    {\cal{I}}_2(k,r) & = &  \int{e^{ikz} \left(\left.\frac{\partial h_{zz}}{\partial t}\right|_{t=0} + \left.\frac{\partial h_{\phi\phi}}{\partial t}\right|_{t=0}\right) dz}
\end{eqnarray}
we have,
\begin{eqnarray}\label{ivp10}
   {\cal{I}}_1(k,r) & = & 2 \pi\sum_j{\left(C^{(+)}_j(k)+C^{(-)}_j(k) \right) G_1^{(j)}(k,r)} \nonumber \\
    {\cal{I}}_2(k,r) & = &   2 \pi i\sum_j{\Omega_j(k)\left(C^{(+)}_j(k)-C^{(-)}_j(k) \right) G_1^{(j)}(k,r)}
\end{eqnarray}

The functions $G_1^{(j)}(k,r)$ do not satisfy orthonormality or other immediately useful conditions. Nevertheless, if we go back to (\ref{gs40}), with $C_{G_1}=0$, we have,
\begin{equation}\label{ivp12}
    H_3^{(j)}(|k|^{1/(1+\kappa)^2}r)= \frac{k^2r^{2(1+\kappa)^2}+\kappa(2+\kappa)}{\sqrt{r}r^{2\kappa+\kappa^2}}
    \frac{\partial}{\partial r}\left(\frac{r^{2+2\kappa+\kappa^2}}{k^2r^{2(1+\kappa)^2}+\kappa(2+\kappa)} G_1^{(j)}(k,r)\right)
\end{equation}
Therefore, if we define,
\begin{eqnarray}\label{ivp014}
   {\cal{J}}_1(k,r) & = &  \frac{k^2r^{2(1+\kappa)^2}+\kappa(2+\kappa)}{\sqrt{r}r^{2\kappa+\kappa^2}}
    \frac{\partial}{\partial r}\left(\frac{r^{2+2\kappa+\kappa^2}}{k^2r^{2(1+\kappa)^2}+\kappa(2+\kappa)} {\cal{I}}_1(k,r)\right)  \nonumber \\
    {\cal{J}}_2(k,r) & = &  \frac{k^2r^{2(1+\kappa)^2}+\kappa(2+\kappa)}{\sqrt{r}r^{2\kappa+\kappa^2}}
    \frac{\partial}{\partial r}\left(\frac{r^{2+2\kappa+\kappa^2}}{k^2r^{2(1+\kappa)^2}+\kappa(2+\kappa)} {\cal{I}}_2(k,r)\right)\;,
\end{eqnarray}
we have,
\begin{eqnarray}\label{ivp16}
   {\cal{J}}_1(k,r) & = & 2 \pi\sum_j{\left(C^{(+)}_j(k)+C^{(-)}_j(k) \right) H_3^{(j)}(|k|^{1/(1+\kappa)^2}r)} \nonumber \\
    {\cal{J}}_2(k,r) & = &   2 \pi i\sum_j{\Omega_j(k)\left(C^{(+)}_j(k)-C^{(-)}_j(k) \right) H_3^{(j)}(|k|^{1/(1+\kappa)^2}r)}
\end{eqnarray}
and we may use the orthonormality of the $H_3^{(j)}(x)$ to obtain,
\begin{eqnarray}\label{ivp18}
   \int_0^{\infty}{H_3^{(\ell)}(|k|^{1/(1+\kappa)^2}r){\cal{J}}_1(k,r) dr}& = & \frac{2 \pi}{k^{1/(1+\kappa)^2}}\left(C^{(+)}_{\ell}(k)+C^{(-)}_{\ell}(k) \right)  \nonumber \\
    \int_0^{\infty}{H_3^{(\ell)}(|k|^{1/(1+\kappa)^2}r) {\cal{J}}_2(k,r) dr} & = &   \frac{2 \pi}{k^{1/(1+\kappa)^2}}\Omega_{\ell}(k)\left(C^{(+)}_{\ell}(k)-C^{(-)}_{\ell}(k) \right)  \;,
\end{eqnarray}
which can be used to solve for the coefficients $C^{(\pm)}_{\ell}(k)$. Therefore, we may invert (\ref{ivp02}) and, given $h_{zz}$ and $h_{\phi\phi}$, obtain the corresponding expansion coefficients. In fact, we only need $h_{zz}(r,0,z)$, $h_{\phi\phi}(r,0,z)$, $(\partial h_{zz}(r,t,z)/\partial t)|_{t=0}$ and $(\partial h_{\phi \phi}(r,t,z)/\partial t)|_{t=0}$. This immediately suggests that, given {\em any} solution $h_{ab}(r,t,z)$ of the perturbation equations, or, rather, of the initial value problem, we may use (\ref{ivp18}) to compute the coefficients $C^{(\pm)}_{\ell}(k)$ and find the evolution of the system that results from that initial data. Although this is conceptually correct, it requires closer examination. If we go back to (\ref{ivp014}), we notice that the right hand sides act as projectors, since any portion of the initial data such that,
\begin{eqnarray}
\label{ivp20}
   {\cal{I}}_1(k,r)  &=& q_1 \frac {k^2r^{2(1+\kappa)^2}+\kappa(2+\kappa)} {r^{2+2\kappa+\kappa^2}} \nonumber \\
  {\cal{I}}_2(k,r)  &=& q_2 \frac {k^2r^{2(1+\kappa)^2}+\kappa(2+\kappa)} {r^{2+2\kappa+\kappa^2}}
\end{eqnarray}
where $q_1$ and $q_2$ are constants, will be deleted from the computation of the coefficients $C^{(\pm)}_{\ell}(k) $. But we have already notice that this type of functional dependence is pure gauge and can be eliminated by an appropriate coordinate transformation. We may understand this feature by noticing that for a single mode, the gauge invariant $\widetilde{G}_1$ is given by,
\begin{eqnarray}\label{1vp22}
 \widetilde{G}_1(r,t,z) & = &  e^{i(\Omega_j(k)t-kz)}\left( K_1(r)+\frac{\kappa}{1+\kappa}K_2(r) - \frac{ k^2 r^{2(1+\kappa)^2}}{\kappa(1+\kappa)(2+\kappa)} K_2(r) \right) \nonumber \\
 & = &  e^{i(\Omega_j(k)t-kz)}\frac{\left(\kappa(2+\kappa)+k^2 r^{2(1+\kappa)^2}\right)^2}{2\kappa(2+\kappa)(1+\kappa)^2}\frac{d}{dr}\left(\frac{r^{2+2\kappa+\kappa^2}}
 {\kappa(2+\kappa)+k^2 r^{2(1+\kappa)^2}} G_1(r)\right) \\
 & = &  e^{i(\Omega_j(k)t-kz)}\frac{\left(\kappa(2+\kappa)+k^2 r^{2(1+\kappa)}\right)^2}{2\kappa(2+\kappa)(1+\kappa)^2 \sqrt{r}} H_3^{(j)}(|k|^{1/(1+\kappa)^2} r) \nonumber
\end{eqnarray}
and, therefore, the functions $H_3^{(j)}(|k|^{1/(1+\kappa)^2} r)$ are themselves gauge invariant, and an expansion of the form (\ref{ivp02}), even if obtained from arbitrary initial data, evolves only the gauge non trivial part of the perturbation. In particular, any initial data that has non vanishing projection on $H_3^{(0)}(|k|^{1/(1+\kappa)^2} r)$ will lead to a gauge non trivial unstable evolution, diverging exponentially in time.

\section{Final Comments}

In this paper we analyzed the axial gravitational perturbations of an infinite line source, and, after imposing boundary conditions at the symmetry axis and at radial infinity, such that the perturbations, in the sense of initial data, can be made arbitrarily smaller than the background, we found the general solution of the corresponding linearized Einstein equations. We analyzed also the problem of the gauge invariance of the solutions, and found a complete set of gauge non trivial solutions, with which we can describe the evolution of arbitrary initial data for the perturbations. The main result of our analysis is that the evolution will contain generically unstable components, and, therefore, that the space time considered here is gravitationally unstable. Since this space time contains a naked singularity, one would be tempted to ascribe the instability to the presence of that singularity. We remark, however, that here, as in the cases considered in \cite{Schw}, and \cite{RN}, the unstable mode is related to the form of a ``potential'' away from the singularity, indicating the possibility that the instability might remain even after smoothing the (curvature) singularity by considering an extended source. In the present case, we would have to consider a cylindrical regular source, such as an infinite cylinder of some kind of matter. The problem in this construction is that the resulting system is considerably more complex than the vacuum case considered here, since, asides from  conditions that must be imposed at the matter - vacuum boundary, we must incorporate an equation of state for the matter, and the resulting appropriate boundary conditions. The question, nevertheless, is interesting, but outside the scope of the present research.

\section*{Acknowledgments}

This work was supported in part by CONICET (Argentina). I am grateful to Gustavo Dotti for his helpful suggestions and criticisms.

 \end{document}